\documentclass[12pt]{article}
\usepackage{amsmath,amssymb,mathrsfs,url,graphicx}

\title{Bohmian Mechanics}
\author{
Roderich Tumulka\footnote{Fachbereich Mathematik, Eberhard-Karls-Universit\"at,
	Auf der Morgenstelle 10, 72076 T\"ubingen, Germany. E-mail: roderich.tumulka@uni-tuebingen.de}
}
\date{August 15, 2019}

\newcommand{\be}{\begin{equation}}
\newcommand{\ee}{\end{equation}}
\newcommand{\foliation}{\mathscr{F}}

\renewcommand{\Im}{\mathrm{Im}}
\newcommand{\PPP}{\mathbb{P}}
\newcommand{\RRR}{\mathbb{R}}
\newcommand{\CCC}{\mathbb{C}}
\newcommand{\scp}[2]{\langle #1|#2 \rangle}
\newcommand{\vQ}{\boldsymbol{Q}}
\newcommand{\vV}{\boldsymbol{V}}
\newcommand{\vj}{\boldsymbol{j}}
\newcommand{\vq}{\boldsymbol{q}}

\begin{document}
\maketitle
\begin{abstract}
Bohmian mechanics, also known as pilot-wave theory or de Broglie--Bohm theory, is a formulation of quantum mechanics whose fundamental axioms are not about what observers will see if they perform an experiment but about what happens in reality. It is therefore called a ``quantum theory without observers,'' alongside with collapse theories \cite{Bell87,Lew} and many-worlds theories \cite{Sau,AGTZ11} and in contrast to orthodox quantum mechanics. It follows from these axioms that in a universe governed by Bohmian mechanics, observers will see outcomes with exactly the probabilities specified by the usual rules of quantum mechanics for empirical predictions. Specifically, Bohmian mechanics asserts that electrons and other elementary particles have a definite position at every time and move according to an equation of motion that is one of the fundamental laws of the theory and involves a wave function that evolves according to the usual Schr\"odinger equation. Bohmian mechanics is named after David Bohm (1917--1992), who was, although not the first to consider this theory, the first to realize (in 1952) that it actually makes correct predictions.
\end{abstract}

\section{Fundamental Laws of Bohmian Mechanics}

While Bohmian mechanics has been considered as a tool for visualization \cite{philippidis79}, for the efficient numerical simulation of the Schr\"odinger equation \cite{Chat}, and other applications \cite{Ori}, the main interest in it arises from the fact that Bohmian mechanics provides a possible way how our world might be and might work.

In its non-relativistic form, the theory asserts the following: $N$ material points (``particles'') move in 3-dimensional Euclidean space (denoted for simplicity as $\RRR^3$) in a way governed by a field-like entity that is mathematically given by a wave function $\psi$ (as familiar from standard quantum mechanics). More precisely,
the position $\vQ_k(t)$ of particle number $k$ at time $t$ obeys Bohm's equation of motion
\be\label{Bohm}
\frac{d\vQ_k(t)}{dt} = \frac{\hbar}{m_k} \Im \frac{\psi^*\nabla_k \psi}{\psi^*\psi}(Q(t),t)\,,
\ee
where $Q(t)=(\vQ_1(t),\ldots,\vQ_N(t))\in\RRR^{3N}$ denotes the configuration of the particle system at time $t$, $m_k$ is the mass of particle $k$, Im the imaginary part of a complex number, $\psi:\RRR^{3N+1}\to \CCC^d$ is the wave function, and $\psi^*\phi$ denotes the scalar product in spin space $\CCC^d$,
\be
\psi^*\phi = \sum_{s=1}^d \psi_s^* \, \phi_s\,.
\ee
For spinless particles ($d=1$), $\psi^*$ is the usual complex conjugate of $\psi$, and the factor $\psi^*$ cancels out of \eqref{Bohm}; in that case, we can express $\psi$ in terms of its modulus $R$ and phase $S$ as $\psi=R\,e^{iS/\hbar}$, and obtain $m_k^{-1}\nabla_k S$ for the right-hand side of \eqref{Bohm}. Yet another way of writing the right-hand side of \eqref{Bohm} is $\vj_k/\rho$, where $\vj_k=(\hbar/m_k) \Im (\psi^*\nabla_k \psi)$ is the quantity known in quantum mechanics as the probability current and $\rho=\psi^*\psi$ the probability density.

The wave function $\psi(q,t)=\psi(\vq_1,\ldots,\vq_N,t)$ evolves with time $t$ according to the usual Schr\"odinger equation
\be\label{Schr}
i\hbar \frac{\partial \psi(q,t)}{\partial t} = -\sum_{k=1}^N \frac{\hbar^2}{2m_k} \nabla_k^2 \psi(q,t) + V(q) \psi(q,t)\,,
\ee
where $V:\RRR^{3N}\to \RRR$ is the potential function, for example the Coulomb potential
\be
V(\vq_1,\ldots,\vq_N) = \frac{1}{2} \sum_{j\neq k} \frac{e_j e_k}{|\vq_j-\vq_k|}
\ee
with $e_j$ the electric charge of particle $j$. The state of the system at time $t$ is described by the pair $\bigl(Q(t),\psi(t)\bigr)$, and Equations \eqref{Bohm} and \eqref{Schr} together determine the state at any other time; thus, Bohmian mechanics is a deterministic theory.

By a mathematical fact known as the ``equivariance theorem'' \cite{Bohm52,DGZ92}, if the initial configuration $Q(0)$ is random with probability density $|\psi(q,t=0)|^2$, then at every time $t$,
\be\label{Born}
Q(t)\text{ is }|\psi(q,t)|^2\text{ distributed}.
\ee
It is another basic rule of the theory that the configuration is indeed so distributed---this is the equivalent of the Born rule in Bohmian mechanics. As discussed in Section~\ref{sec:typicality} below, this rule follows if it is assumed that the initial configuration of the universe is typical relative to the $|\Psi|^2$ distribution, where $\Psi$ is the initial wave function of the universe.

We note that Bohmian mechanics is time reversal invariant: if $t\mapsto \bigl(Q(t),\psi(t)\bigr)$ is a solution of \eqref{Bohm} and \eqref{Schr}, then so is $t\mapsto \bigl(Q(-t),\psi^*(-t)\bigr)$, and also Born's rule still holds for the time reverse. Likewise, Bohmian mechanics is invariant under Galilean boosts, spatial rotations, and space-time translations (see, e.g., \cite{DGZ92}). I also note that equations of motion analogous to \eqref{Bohm} have been devised for a wide class of Hamiltonians \cite{SV09}, not just non-relativistic Schr\"odinger operators as in \eqref{Schr}. In particular, the equation of motion for the 1-particle Dirac equation \cite{Bohm53,BH} asserts that the possible world-lines (i.e., particle paths in space-time) are the integral curves of the probability current 4-vector field $j^\mu=\overline{\psi}\gamma^\mu \psi$.

\section{Example: The Double-Slit Experiment}

\begin{figure}[h]
\begin{center} 
  \includegraphics[width=0.8\textwidth]{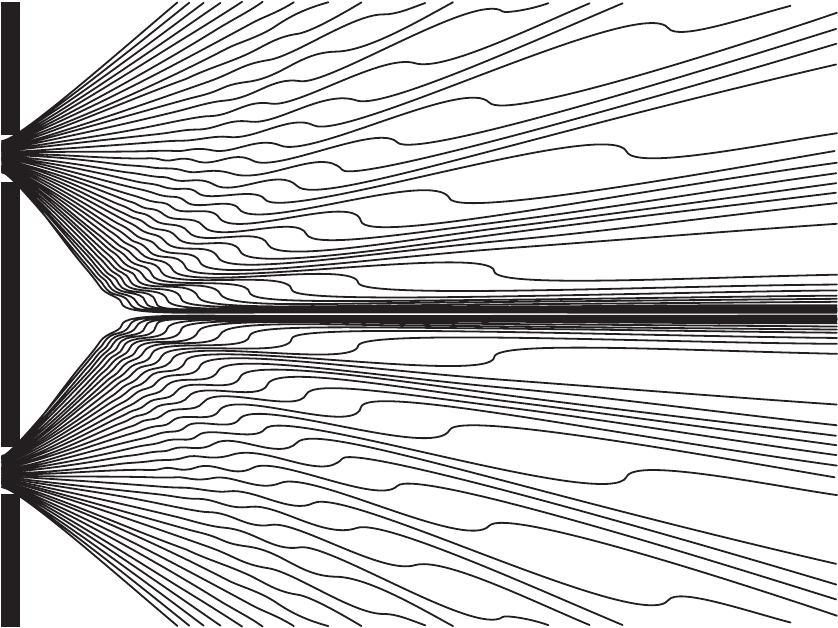}
\end{center}
 \caption{Several possible trajectories for a Bohmian particle in a double-slit setup, coming from the left. (Reprinted from \cite{DT}, based on a figure in \cite{philippidis79}.)}\label{pic}
\end{figure}

In Bohmian mechanics there is wave--particle duality in a literal sense: There is a wave, and there is a particle. In the double-slit experiment, the wave passes through both slits, whereas the particle passes through only one slit in each run. In Figure~\ref{pic}, a sample of 80 alternative trajectories is shown.

We may place a detecting screen at the right end of the figure. Since the trajectories in the sample are roughly $|\psi|^2$ distributed, their arrival points are also $|\psi|^2$ distributed by equivariance. One can see in Figure~\ref{pic} that there are some locations in which more trajectories arrive---in fact, those are the locations where $|\psi|^2$ is large; thus, an interference pattern with bright and dark fringes is visible in the arrival points on the screen, in agreement with the empirical findings. This is how Bohmian mechanics explains the double-slit experiment, without any paradox remaining \cite{Bell80}.

For the outcome of the experiment, it is crucial that the motion of the Bohmian particle is non-Newtonian, and that its trajectory can be bent also in the absence of external force fields. It is bent because the equation of motion \eqref{Bohm} dictates it. Note also that the motion of the particle, if it went through the upper slit, will depend also on the part of the wave function that went through the lower slit. In particular, the motion of the particle depends on whether both slits are open, and would have been different if one slit had been closed. With only one slit open, the distribution of the arrival points on the screen would have created a different interference pattern. If both slits are open, then the particle, although it passes only through one slit, ``knows'' this because the wave passed through both slits. Note that the Bohmian explanation of the double-slit experiment involves no faster-than-light effects and no retrocausation. This is still true in Wheeler's \cite{delayedchoice} ``delayed choice'' variant of the experiment, which was explained using Bohmian mechanics in \cite{Bell80}.

\section{Empirical Predictions}
\label{sec:predictions}

An analysis of Bohmian mechanics shows that its empirically testable predictions agree exactly with those of standard quantum mechanics, whenever the latter are unambiguous \cite{Bohm52,Bell80,DGZ92}. Thus, Bohmian mechanics is a counter-example to the claim put forward by Niels Bohr \cite{Bohr} (and often repeated since) that in quantum mechanics a single coherent picture of reality be impossible. In particular, it turns out that the statistics of outcomes of experiments are related to the operators known as ``observables'' in the same way as in standard quantum mechanics. The fact that these operators in general do not commute has often in orthodox quantum mechanics been taken as a sign that a ``realistic'' picture be impossible, whereas Bohmian mechanics shows how it is, in fact, possible; see Section~\ref{sec:observables} below for more detail.

The predictive rules of standard quantum mechanics include the collapse of the wave function; in fact, an effective collapse of the wave function of a system $s$ comes out of the equations of Bohmian mechanics, although the wave function of the universe never collapses. Here is how. In a (hypothetical) universe governed by Bohmian mechanics, also observers and measurement apparatus are made of Bohmian particles, and are jointly governed by Equations \eqref{Bohm} and \eqref{Schr}. During the experiment, the initial wave function $\psi_0(q_s,q_a)= \varphi(q_s)\,\chi(q_a)$ of the system $s$ and apparatus $a$ together evolves according to \eqref{Schr} to $\psi_t(q_s,q_a)$. Suppose that whenever $\varphi$ is an eigenfunction $\varphi_\alpha$ of the self-adjoint operator $A$ with eigenvalue $\alpha$, the final wave function is of the form
\be\label{eigenfunction}
\psi_t(q_s,q_a)=\varphi_\alpha(q_s)\, \chi_\alpha(q_a)\,,
\ee
where $t$ is the duration of the experiment, and $\chi_\alpha$ is a wave function of an apparatus displaying the outcome $\alpha$; this would be the case in an ``ideal quantum measurement'' of $A$. Then it follows by the linearity of \eqref{Schr} for an arbitrary superposition $\varphi=\sum_\alpha c_\alpha \, \varphi_\alpha$ of eigenfunctions that
\be\label{postmeasurement}
\psi_t(q_s,q_a) = \sum_\alpha c_\alpha\, \varphi_\alpha(q_s) \, \chi_\alpha(q_a)\,. 
\ee
Since the actual configuration $Q(t)=(Q_s,Q_a)$ is $|\psi_t|^2$ distributed, and since $\chi_\alpha$ is concentrated on those configurations in which the apparatus displays the outcome $\alpha$ (say, by the position of a pointer), $Q_a$ has probability $|c_\alpha|^2$ to be such a configuration; this probability value for obtaining the outcome $\alpha$ agrees with the rules of standard quantum mechanics. Moreover, since the various $\chi_\alpha$ have macroscopically disjoint supports in configuration space, the packets $\psi_\alpha(q_s,q_a)=\varphi_\alpha(q_s)\, \chi_\alpha(q_a)$ will remain non-overlapping for the next $10^{100}$ years or more. But as long as $\psi=\sum_\alpha c_\alpha \,\psi_\alpha$ consists of non-overlapping packets, the motion of $Q(t)$, which is governed according to \eqref{Bohm} by $\psi$ and its derivatives at $Q(t)$, will depend only on one packet, the one containing $Q(t)$. Thus, it makes no difference for the future motion of all particles (at least for the next $10^{100}$ years) if we drop all other packets $\psi_{\alpha'}$ and replace $\psi$ by $\psi_\alpha$ with $\alpha$ the actual outcome. The $|\psi_\alpha|^2$ distribution represents the conditional probability distribution of $Q(t)$, given that $Q(t)$ lies in the support of $\psi_\alpha$ (i.e., given that the outcome was $\alpha$). That is why the wave function can be collapsed ($\psi\mapsto\psi_\alpha$), yielding an eigenfunction $\varphi_\alpha$ for the system $s$. I called it an ``effective'' collapse because one can practically replace $\psi\mapsto \psi_\alpha$ although the true wave function is still the uncollapsed $\psi$.

More generally (beyond ideal quantum measurements), in (spinless) Bohmian mechanics every system $s$ can be attributed a wave function of its own, the \emph{conditional wave function}
\be\label{psiconddef}
\psi^{(s)}(q_s) = \psi(q_s,Q_a)
\ee
obtained by inserting the actual configuration of the environment $a$ (and normalizing if desired). (In the presence of spin, one uses a conditional density matrix \cite{density}.) In the situation discussed before, $\psi^{(s)}\propto \varphi$ before the measurement and $\psi^{(s)}\propto \varphi_\alpha$ afterwards. The conditional wave function evolves according to the Schr\"odinger equation \eqref{Schr} if there is no interaction between $s$ and $a$ and $s$ is suitably disentangled from $a$ (viz., $\psi(q_s,q_a)=\varphi(q_s) \, \chi(q_a)+\psi^\perp(q_s,q_a)$ with the $q_a$-support of $\psi^{\perp}$ macroscopically disjoint from $Q_a$), but not in general. In a measurement-like situation, the conditional wave function undergoes a genuine collapse.

In orthodox quantum mechanics (OQM), one cannot form the conditional wave function as there is no $Q_a$. What is worse, since OQM insists that there are no variables (such as $Q$) in addition to the wave function, and since the post-measurement wave function \eqref{postmeasurement} is a superposition of terms corresponding to all possible outcomes, there is nothing in the state according to OQM at time $t$ that would represent the actual outcome---a problem known as the quantum measurement problem. The only ways out are (i)~to introduce further variables (as in Bohmian mechanics), (ii)~to deny the linearity of the time evolution of the wave function (as in collapse theories such as GRW \cite{Bell87,Lew}), or (iii)~to deny that there is a single outcome (as in many-worlds theories \cite{Sau,AGTZ11}).

\section{Limitations to Knowledge and Control}

In classical mechanics, one usually pretends that one can measure the state of a system (position and momentum of all particles) to arbitrary accuracy without disturbing it, and that one can in principle prepare a system in any state. In Bohmian mechanics, in contrast, there are sharp limitations to knowledge and control: inhabitants of a Bohmian universe cannot know the position of a particle more precisely than allowed by the $|\psi|^2$ distribution \cite{DGZ92}, where $\psi$ is its conditional wave function, and they cannot prepare a particle with a specific position $\vq\in\RRR^3$ and a wave function $\psi$ other than a Dirac delta function (i.e., a wave function concentrated on a single point). Furthermore, they cannot measure the position at time $t$ without disturbing the particle: its future trajectory will be significantly different from what it would have been without a measurement. A limitation to knowledge that applies to all versions of quantum mechanics is that wave functions cannot be measured (e.g., \cite{CT16}); i.e., if Alice prepares a particle with a wave function $\psi$ of her choice then Bob cannot find out, from experiments on the particle, what $\psi$ is. He could determine $\psi$, however, if Alice prepared many particles, each with the same wave function $\psi$. Further such limitations are described in Sections~\ref{sec:nonlocality} and \ref{sec:observables} below. Some researchers feel that theories that entail limitations to knowledge are bad; since, as just pointed out, every version of quantum mechanics must admit such limitations, such a sentiment seems inadequate. 

Variables besides the wave function, such as the position variables $\vQ_k(t)$ in Bohmian mechanics, are traditionally called ``hidden variables.'' This terminology has stuck but one should be aware that it is rather misleading: $\vQ_k(t)$ can in principle be measured at any time $t$ to any desired accuracy (so it is hardly ``hidden''), whereas $\psi_t$ cannot---so it is really the wave function, which Bohmian mechanics has in common with OQM, which is a ``hidden variable'' \cite{Bell87}.

\section{Differences to Orthodox Quantum Mechanics}

Orthodox quantum mechanics (OQM) is the traditional understanding of quantum mechanics which goes back to the ``Copenhagen interpretation'' of Niels Bohr. It
insists that quantum particles do not have trajectories, and more generally that there are no further variables besides $\psi$ (for microscopic systems), so that $\psi$ is the complete description of the state (e.g., \cite{Bohr}, \cite[Chap.~III.2]{vN}, \cite[p.~5--6]{LL}), whereas in Bohmian mechanics the state (i.e., the real factual situation) is represented by the pair $(Q,\psi)$. On the other hand, it is assumed in OQM that macroscopic (``classical'') quantities do always have sharp values and cannot be in a superposition, so there are also some kind of ``hidden variables'' in OQM, except that it remains vague which quantities exactly are supposed to be ``classical,'' and which equations govern them. In Bohmian mechanics, a categorization of systems as ``classical'' and ``quantum'' may be convenient but is fundamentally unnecessary, as the entire universe is governed by the laws \eqref{Bohm}, \eqref{Schr}, \eqref{Born}, so there is no problem about regarding the entire universe as quantum.

A deeper difference between OQM and Bohmian mechanics concerns a philosophical attitude that can be called ``positivism'' and whose central idea is that a statement is unscientific or even meaningless if it cannot be tested experimentally, that an object is not real if it cannot be observed, and that a variable is not well-defined if it cannot be measured. Positivism suggests that science should limit itself to operational statements such as ``if one performs experiment $X$ then one obtains outcome $Y$ with probability $Z$.'' That is why the usual axioms of quantum mechanics are about what observers will see. I regard this form of positivism as exaggerated and implausible: for example, events behind the horizon of a black hole may be unobservable to all outside observers but would very much seem to be as real as outside events. In fact, I regard positivism as \emph{refuted} by the limitations to knowledge such as the impossibility of measuring wave functions. Nevertheless, in the physics literature one often reads arguments with a positivistic flavor, and OQM is particularly positivistically influenced. In contrast, Bohmian mechanics takes a ``realist'' attitude, according to which we should make the best hypothesis we can about what actually happens in reality. The reason we make these hypotheses is that we would like to find out how the world works \cite{Mau16}. 

A related philosophical difference between OQM and Bohmian mechanics is that OQM finds it acceptable to say contradictory things about what happens in reality as long as all operational predictions remain unaffected by the contradictions. This attitude, which makes OQM incoherent from a realist perspective, is often regarded by advocates of OQM as an aspect of complementarity; it is often accompanied by the attitude that everything we say about what happens in reality is anyway only metaphor and should not be taken seriously. In contrast, in Bohmian mechanics there are no contradictions and there is no need for metaphors, as we can easily and clearly say what actually happens in a Bohmian universe. That is because Bohmian mechanics provides a single coherent picture of reality.

A further difference concerns the status of observables. In OQM it is common to talk about observables as if they were quantities, as if they had values. It should not be surprising that paradoxes arise from such loose talk. In Bohmian mechanics, also this problem is absent as we are dealing with variables that actually have values; more about this in Section~\ref{sec:observables} below.

\section{Non-Locality}
\label{sec:nonlocality}

We now turn to Bell's theorem \cite{Bell64,GNTZ,Bell87b,Mau0}. It is sometimes claimed (e.g., \cite[p.~53]{Wig}, \cite{Hawking}) that Bell's theorem excludes hidden-variable theories (such as Bohmian mechanics) because it implies that any hidden-variable theory has to be non-local (i.e., involve faster-than-light influences), and that would be in conflict with relativity. This picture is not quite right. Bell's theorem actually shows that the observed probabilities in certain experiments (Einstein-Podolsky-Rosen-Bell experiments) are incompatible with locality, so our world must be non-local, and every theory in agreement with experiment must be non-local. Since these observed probabilities are predicted by the predictive rules of quantum theory, one can also conclude that every theory in agreement with these rules must be non-local. Bohmian mechanics agrees with these rules, and it is non-local. Thus, Bell's theorem does not at all exclude Bohmian mechanics but, on the contrary, proves its non-local character inevitable. And non-locality need not conflict with relativity, as illustrated particularly by the relativistic Ghirardi--Rimini--Weber (GRW) theory \cite{Tum06,Mau0}.

The widespread misperception that Bell's theorem concern only hidden-variable theories and have nothing to say about other approaches presumably originates from the fact that John Bell's original 1964 paper \cite{Bell64} cited the famous 1935 paper of Einstein, Podolsky, and Rosen (EPR, \cite{EPR}) for showing (as it did) that locality implies the existence of hidden variables, so that Bell focused on excluding the remaining possibility of local hidden-variable theories in order to exclude all local theories. EPR's argument was and is often not sufficiently appreciated, and its relevance to Bell's argument often missed (see \cite{GNTZ,Mau1} for elaboration).

The non-locality shows up in Bohmian mechanics in the dependence of the velocity of particle 1 on the position of particle 2 according to \eqref{Bohm}; since this dependence is instantaneous, no matter how big the distance between the two particles, it is a faster-than-light influence. As a consequence, for devising a version of Bohmian mechanics in a relativistic space-time \cite{HBD}, we need a temporal order between spacelike-separated space-time points, or a notion of simultaneity-at-a-distance. Such a notion can be mathematically expressed by a foliation (slicing) $\foliation$ of space-time into disjoint spacelike hypersurfaces (sometimes called ``the time foliation''). In such versions of Bohmian mechanics that have been studied, it turns out that the probabilities of experimental outcomes do not depend on $\foliation$, so that inhabitants of such a universe cannot find out empirically which foliation $\foliation$ is. So in this theory, there is another limitation to knowledge: $\foliation$ cannot be determined experimentally. While positivistically inclined researchers find this objectionable, I find positivism objectionable. In contrast to the ether, which could simply be eliminated from classical electrodynamics, $\foliation$ cannot be eliminated from Bohmian mechanics without a breakdown of the equations. This situation suggests that $\foliation$ plays a legitimate role in the theory. The theory could still be regarded as relativistic if $\foliation$ is determined by a covariant law; proposals of such laws are outlined in \cite{Tum07,DGNSZ14}.

\section{Observables}
\label{sec:observables}

An analysis of the fundamental laws of Bohmian mechanics shows that there is no apparatus with which the inhabitants could measure the velocity of a particle without any prior knowledge of its wave function \cite{DGZ04}; this is another limitation to knowledge. In contrast, if the wave function is known, or is known to belong to a suitable class of wave functions (e.g., to be in the classical regime), then the velocity can be measured; in particular, there is no obstacle to measuring the velocity of a macroscopic object in the classical regime. Likewise, if many particles are given, and we know that each one has the same wave function, then the velocity of (almost) each one can be measured (as we then can determine the wave function). A specific procedure \cite{Wis07,DGZ09} of this kind, so-called weak measurements of the velocities, has been carried out experimentally \cite{Steinberg}, resulting in a picture similar to Figure~\ref{pic}. 

Furthermore, if we ask about the asymptotic velocity $u$ as $t\to\infty$ (that the particle will ultimately have if from now on no external forces act on it) instead of the instantaneous velocity $v$ at time $t$, then the situation is different: $u$, which is determined by $Q$ and $\psi$, can be measured without knowledge of $\psi$, and the product $mu$ (mass times asymptotic velocity) has the distribution known in standard quantum mechanics as the momentum distribution, i.e., (leaving out factors of $\hbar$) $|\hat\psi(k)|^2$ with $\hat\psi(k)$ the Fourier transform of $\psi(q)$. The Heisenberg uncertainty relation then means the following in Bohmian mechanics: While position and momentum (understood as $mu$) of a particle do have actual values, inhabitants cannot know both values with inaccuracies whose product is smaller than $\hbar/2$, even if they know the particle's wave function. (This is another limitation to knowledge.)

In contrast to position and momentum ($mu$), energy, angular momentum, and spin do not even have actual values (except for special wave functions, viz., eigenfunctions of energy or angular momentum or spin). Rather, the experiments that are commonly called ``quantum measurements'' of energy etc., create random outcomes instead of revealing pre-existing quantities (as in the ordinary meaning of ``measuring''). 
Thus, ironically, Bohmian mechanics is a ``no hidden variables'' theory for most observables. 

(The outcomes are, in fact, determined by the initial wave function $\chi(q_a)$ of the apparatus, its initial configuration $Q_a$, the coupling Hamiltonian, and the object's initial wave function $\varphi(q_s)$ and configuration $Q_s$ together; but they are in general not functions of $\varphi$ and $Q_s$ alone. See also \cite{Emery} for a discussion of deterministic randomness.)

For example, a Stern--Gerlach experiment, the usual experiment for a quantum measurement of a component of spin, say $\sigma_z$, uses a magnetic field to direct the up-component of $\psi$ into a different spatial region than the down-component, and then uses a detector to find out in which region the particle is. When the particle is found in the region containing the up-component, one says that the outcome is ``spin up.'' This example shows several things: First, that there is no need to introduce an actual value for spin components, as the Stern--Gerlach experiment is explained with positions and wave functions alone. Second, that $z$-spin then is not a well-defined quantity in Bohmian mechanics, except when the wave function is an eigenstate of the Pauli matrix $\sigma_z$; the value of $z$-spin is only created during the experiment. Third, that there can be different experiments, all of which are ``quantum measurements'' of the same observable $\sigma_z$, but which lead to different outcomes when applied to particles in the same Bohmian initial state $(Q,\psi)$: indeed, inverting the field and interchanging the regions for ``up'' and ``down'' would still lead to the same statistics of ``up'' and ``down'' for a given $\psi$ but not to the same outcome for fixed $Q$ and $\psi$. (Concretely, the inverted field would deflect the $z$-down component of $\psi$ upwards, and in certain simple cases with $\psi=|x\text{-up}\rangle$, the Bohmian particle would be deflected upwards whenever the initial position is in the upper half of the packet, resulting in outcome ``down'' with the inverted field but outcome ``up'' with the original field; see, e.g., \cite[Sec.~V]{Nor14} for details.) And fourth, it further follows that it is, in fact, not possible to introduce an actual value $S_z$ of $\sigma_z$ such that every Stern--Gerlach experiment that qualifies as a quantum measurement of $\sigma_z$ yields $S_z$ as the outcome. See \cite{Nor14} for elaboration about spin in Bohmian mechanics.

There are various ``no hidden variables'' theorems \cite{Bell66,HS11}. Some of them, such as the Kochen--Specker theorem, do not apply to Bohmian mechanics because Bohmian mechanics does not associate actual values with most quantum observables. However, for other theorems, such as the Bell inequality theorem, this does not matter much because we may choose a specific experiment (including, if necessary, the wave function and configuration of the apparatus) for each of the relevant observables, and Bohmian mechanics then does determine the outcome as a function of the particle's $Q$ and $\psi$. Specifically, the Bell inequality theorem then shows that for an EPR pair ($Q=(\vQ_a,\vQ_b)$, $\psi=2^{-1/2} (|\!\uparrow\downarrow\rangle - |\!\downarrow\uparrow\rangle) \, \phi_a(\vq_a)\, \phi_b(\vq_b)$, where the supports of $\phi_a$ and $\phi_b$ are separated) on which Alice and Bob each carry out a Stern--Gerlach experiment about a spin component of their choice, say Alice first and Bob later, the function that yields Bob's outcome as a function of $Q$ and $\psi$ depends also on Alice's choice. That is, the theorem proves that Bohmian mechanics is non-local, which was clear already before Bell \cite[p.~186]{Bohm52}.

Let us return to a fact mentioned before: Two different experiments, which have different outcomes when acting on the same Bohmian state $(Q,\psi)$, may still have the same \emph{statistics} of outcomes when fed with a random $Q$ that is $|\psi|^2$ distributed. The statistics can, in fact, be expressed in terms of operators, more precisely of a POVM (positive-operator-valued measure) \cite{POVM}; this means in our case that with every possible outcome $z$ of the experiment there is associated a positive self-adjoint operator $F(z)$ such that, when the experiment is applied to a system with wave function $\psi$ (with $\|\psi\|=1$), then the probability distribution of the random outcome $Z$ is given by
\be
\PPP_{\psi}(Z=z) = \scp{\psi}{F(z)|\psi}
\ee
for every $z$ \cite{DGZ04}. That is how operators come up as ``observables'' in Bohmian mechanics: They encode the probability distribution of $Z$ as a function of $\psi$. The operators $F(z)$ have the property that they add up to the identity operator $I$, $\sum_z F(z) = I$. In the special case that the $z$ are real numbers and the $F(z)$ are projections, the POVM $F(\cdot)$ can equivalently be represented by the self-adjoint operator
\be
A=\sum_z z\, F(z)
\ee
of which it is the spectral decomposition (i.e., the $z$ are the eigenvalues of $A$, and $F(z)$ is the projection to the eigenspace). That is why self-adjoint operators represent observables in many relevant cases.

So two different experiments can have equal statistics for every $\psi$, and this happens when they have the same POVM $F(\cdot)$. Since in orthodox terminology one says in this case that the two experiments ``measure the same quantum observable,'' it can be difficult to appreciate that they can have different outcomes (in particular because in OQM one cannot ask the question whether they have different outcomes). For example, opponents of Bohmian mechanics have described an experiment whose POVM is the usual position observable but whose outcome is not the position of the particle, have claimed that position measurements disagree with the Bohmian position, and have called the Bohmian trajectories ``surrealistic'' \cite{surr1,surr2}. In fact, this example only illustrates how misleading talk about ``measurements'' of ``observables'' can be: while ordinary particle detectors do find the particles at their Bohmian positions, this particular experiment does not measure position although it is a ``quantum measurement'' of the ``position observable,'' meaning merely that the distribution of its outcomes agrees with $|\psi(q)|^2$.

(Here is another, simpler exampe of an experiment with this property. Consider two normalized 1-dimensional wave packets $\psi_1,\psi_2$ centered around the locations $x_1,x_2$ respectively, with $x_1<x_2$ and $\psi_1$ moving to the right and $\psi_2$ to the left, such that they exchange places in one time unit. On the 2-dimensional subspace $S$ of the superpositions $\psi = c_1 \psi_1 + c_2 \psi_2$ consider, as the analog of the position observable, the operator $A= x_1 |\psi_1\rangle \langle \psi_1| + x_2 |\psi_2\rangle \langle \psi_2|$, which is a coarse-grained position observable. Consider two experiments on a system with normalized initial wave function $\psi\in S$: (i)~measure the (coarse-grained) position at time 0; (ii)~let the system evolve to time 1, then measure the (coarse-grained) position and exchange $x_1 \leftrightarrow x_2$ in the result. Both experiments yield $x_1$ with probability $|c_1|^2$ and $x_2$ with probability $|c_2|^2$, so they are quantum measurements of the same observable ($A$ at time 0). However, since Bohmian trajectories cannot cross, some of them start at time 0 near $x_1$ and end up at time 1 near $x_1$, and in that case the result of experiment (ii) does not agree with the position of the Bohmian particle at time $0$. Note that there is nothing mysterious about the fact that the two experiments have the same distribution although they produce different outcomes.)

\section{History}

That quantum particles may have trajectories ``guided'' by a wave was proposed as early as 1923 by Albert Einstein \cite[p.~463]{WignerEinst} and John Slater \cite[p.~9]{Sla75}, \cite[p.~544]{Mehra} but not published. The equation of motion \eqref{Bohm} was considered, without spin, by Louis de Broglie in 1926 \cite{deB} but later abandoned because he incorrectly believed that the theory made empirically wrong predictions. It was rediscovered independently by Nathan Rosen in 1945 \cite{Ros45} and David Bohm in 1952 \cite{Bohm52}. Bohm was the first to realize that the theory actually makes empirically correct predictions. The correct extension to particles with spin was given by John Bell in 1966 \cite{Bell66}; Bell also contributed to the exploration and clarification of the theory \cite{Bell87b}.

Bohm \cite{Bohm52} formulated the theory in an awkward way by introducing a second-order equation (the one obtained from \eqref{Bohm} by taking a time derivative on both sides and expressing $\partial\psi/\partial t$ via the Schr\"odinger equation) as the equation of motion and demanding that the initial positions $\vQ_j(0)$ and the initial velocities $\vV_j(0)=d\vQ_j/dt$ be related according to a ``constraint condition'' identical to \eqref{Bohm}. Since it then follows mathematically that \eqref{Bohm} is satisfied at all times, Bohm's prescription is actually equivalent to saying that $Q(t)$ is a solution of \eqref{Bohm} at all times. Bohm also suggested introducing additional hidden variables for spin \cite{BH}, an unnecessary move, as discussed above.

\section{Typicality and the Origin of Randomness}
\label{sec:typicality}

Bohm realized, by means analogous to the considerations in Section~\ref{sec:predictions} above, that inhabitants of a universe governed by Bohmian mechanics will observe statistics of outcomes of experiments in agreement with the predictive rules of standard quantum mechanics (for short, ``the quantum statistics''), including the joint distribution of several experiments. A deeper analysis was provided by Detlef D\"urr, Sheldon Goldstein, and Nino Zangh\`\i\  \cite{DGZ92} with a result that parallels the law of large numbers in probability theory as follows. Recall that the outcomes of all experiments are determined by the initial wave function $\Psi=\Psi(0)$ and the initial configuration $Q=Q(0)$ of the universe (even if the necessary calculation cannot be carried out in practice). The result of D\"urr et al.\ shows that, for given $\Psi$, most $Q$ are such that inhabitants throughout the world history will observe the quantum statistics.  Here, ``most'' means the following: we say that ``most $Q$ have property $X$'' if 
\be
\int_S dq\, |\Psi(q)|^2 \text{ is close to }1 
\ee 
with $S$ the set of all $Q$ with property $X$. That is why it suffices, as a fundamental law of Bohmian mechanics, to demand that $Q$ be \emph{typical} with respect to $|\Psi|^2$. This is close to saying that $Q$ was chosen randomly with distribution $|\Psi|^2$ but a little different. For example, the sequence of digits of $\pi$ (314159265\ldots)~``looks random'' in the long run although it cannot be said to \emph{be random}. Put differently, for the purpose of the statistical looks of the sequence, $\pi$ is a typical number in the interval $[0,4]$. Likewise, for $Q$ it is only relevant that it ``looks as if chosen randomly with $|\Psi|^2$ distribution'' and not whether it \emph{is random} in a fundamental sense; and the ``looks'' refers to the macroscopic history of $Q(t)$. Thus, from the perspective of the inhabitants it is indistinguishable from being random, so the typicality of $Q$ is the origin of randomness in a Bohmian universe.

It has been pointed out \cite{VW} that the motion of the Bohmian configuration is often chaotic (even mixing, to be precise), with the consequence that any initial distribution $\rho\neq |\psi|^2$ on configuration space tends to come closer and closer to $|\psi_t|^2$ over time. It has further been suggested \cite{Bohm52,VW} that this phenomenon may explain the $|\psi|^2$ distribution. However, the real question concerns not so much the configuration space of a system as that of the universe; and there it does not seem as if a deeper insight were gained if somebody could show (which has not been shown) that also (say) a $|\Psi|^4$-typical or a $|\Psi|^0$-typical $Q(0)$ would lead to $|\psi|^2$-statistics for the outcomes of most experiments. While the $|\Psi|^2$ distribution is special since it is equivariant, the $|\Psi|^4$ and $|\Psi|^0$ distribution are just some among many conceivable distributions; in fact, if we found empirically (which we have not) that it was necessary to assume that $Q(0)$ was $|\Psi|^4$-distributed, then it would be a big puzzle needing explanation why it was $|\Psi|^4$ of all distributions instead of the natural, equivariant $|\Psi|^2$.

\section{Identical Particles}

The symmetrization postulate of quantum mechanics applies equally in Bohmian mechanics. It asserts that when particle $i$ and particle $j$ belong to the same particle species, then the wave function is either symmetric or anti-symmetric against the permutation of their coordinates and spin indices,
\be
\psi_{...s_i...s_j...}(...\vq_i...\vq_j...) = \gamma\, \psi_{...s_j...s_i...}(...\vq_j...\vq_i...)
\ee
with $\gamma=-1$ if the species is fermionic and $\gamma=+1$ if bosonic. It has been suggested in the literature (e.g., \cite[p.~227]{LL}) that the reason behind the symmetrization postulate is that there are no trajectories in quantum mechanics and therefore no fact about which particle at time $t_1$ is which particle at time $t_2$; however, Bohmian mechanics shows that the symmetrization postulate is compatible with the existence of trajectories. 

On the other hand, if we take the particles seriously, as we do in Bohmian mechanics, then, for a system of $N$ identical particles, the particles should not be numbered, as there is no fact in the physical world about which particle is particle number 1. Therefore, two configurations that differ only by a permutation, such as $(\vq_3,\vq_1,\vq_2)$ and $(\vq_1,\vq_2,\vq_3)$, should be regarded as two mathematical representations of the same physical configuration. Thus, the space of physical configurations is the space ``${}^N\RRR^3$'' of \emph{unordered} configurations \cite{LM77}, i.e., of permutation classes of $N$-tuples (leaving out for simplicity the $N$-tuples containing repetitions such as $\vq_1=\vq_2$) or of $N$-element subsets of physical 3-space, as opposed to the space $\RRR^{3N}$ of \emph{ordered} configurations. Now Bohm's law of motion \eqref{Bohm} is such that for a wave function that is either symmetric or anti-symmetric, two initial configurations that differ only by a permutation evolve to later configurations at any time $t$ that differ only by a permutation (in fact, the same permutation). Therefore, it defines a curve in ${}^N\RRR^3$, as it should. This would not be so for general wave functions (neither bosonic nor fermionic), and this situation can be regarded as a reason behind the symmetrization postulate \cite{Bac03,topid0}. Since ${}^N\RRR^3$ is a topologically non-trivial space, this approach to explaining the symmetrization postulate, which goes back to Jon M.~Leinaas and Jan Myrheim \cite{LM77}, is known as the ``topological'' approach.

\section{The Classical Limit}

In OQM, some systems have to be ``classical'' in order to even make sense of the theory. In Bohmian mechanics, in contrast, no problem would arise even if no system were in a classical regime. Nevertheless, such a regime actually governs many systems, particularly macroscopic ones.  In OQM the classical limit is problematical because of the unsolved problem of how definite properties (e.g., definite positions) of macroscopic objects arise, given that these objects consist of quantum particles (electrons and quarks) that can be in arbitrary superpositions.
No such problem affects Bohmian mechanics: There, particles have trajectories, and the study of the classical limit concerns merely the clear mathematical question under which conditions these trajectories (or the center-of-mass trajectories of macroscopic bodies) will obey Newton's equation of motion
\be\label{Newton}
m_k\frac{d^2\vQ_k(t)}{dt^2}=-\nabla_k V(Q(t))\,.
\ee
See \cite{cl} for an overview of results about this question. In brief, \eqref{Newton} tends to hold when two mechanisms prevail: decoherence \cite{Cru} and the tendency of wave functions to become locally plane waves.

\section{Quantum Field Theory}

Leaving aside the circumstance that in most quantum field theories the Hamiltonians are mathematically ill defined (and that it is unknown how to define them rigorously \cite{Wallace}), Bohmian mechanics can be extended to quantum field theory in several ways. It has not been settled which Bohm-type model is the most convincing one. The proposed extensions differ in whether they use a \emph{field ontology} or a \emph{particle ontology}. For a field ontology (see \cite{Str07} for an overview), one assumes that instead of a particle configuration, the variable besides $\psi$ is a field configuration (a function on 3-space). A particle ontology, in contrast, means to stick with a particle configuration. Also combinations are conceivable: Bohm \cite{Bohm52} proposed to use a particle ontology for fermions and a field ontology for the bosonic degrees of freedom. It seems not possible to set up a field ontology for fermions. 

With a particle ontology, it seems natural to introduce the possibility of creation and annihilation of particles. Such models have been set up \cite{Bell86,crlet}: for a given Hamiltonian $H$ and configuration operators (a projection-valued measure $P(\cdot)$ on configuration space generalizing the position operators), the model is not unique, but uniquely selected by considerations of naturalness and simplicity. These models are no longer deterministic but instead stochastic, as the creation and annihilation events correspond to jump between sectors of configuration space corresponding to different particle numbers, and every jump towards a higher particle number has to be stochastic, or else $Q(t)$ could not be continuously distributed in the higher sector, so that $|\psi|^2$ could not be equivariant. Such considerations lead to a particular rate (probability per time) for the jump from $q'$ to anywhere in the volume element $dq$, given by \cite{crlet}
\be\label{jumprate}
\sigma^\psi(q'\to dq) = \frac{2}{\hbar} \frac{\Im^+\scp{\psi}{P(dq)HP(dq')|\psi}}{\scp{\psi}{P(dq')|\psi}}
\ee
with $x^+=\max\{x,0\}$. 

It may be noteworthy that although Bohmian mechanics is deterministic, neither Bohm nor Bell nor the authors of \cite{crlet} found or find stochastic theories unacceptable. Rather, the simplest and most convincing theory of non-relativistic quantum mechanics (i.e., Bohmian mechanics) happens to be deterministic, and the simplest and most convincing theory of particle creation happens to be stochastic. Speaking of stochastic theories, I note that a variant of Bohmian mechanics for non-relativistic quantum mechanics with stochastic trajectories (a diffusion process) was proposed by Edward Nelson under the name ``stochastic mechanics'' \cite{Nel85,Gol87}. While Nelson's initial hopes that this theory might somehow get rid of the wave function, and that it might provide a superior explanation of randomness, have not materialized, it does provide a coherent and working theory when understood in the same way as Bohmian mechanics. It has received less attention because it is more complicated than Bohmian mechanics (as it involves stochastic processes) while not being more convincing.

Another open question that comes up when setting up Bohmian versions of quantum field theories is whether the Dirac sea should be taken literally. (The Dirac sea is, roughly speaking, the picture that electrons cannot occur in a state of negative energy because all negative energy states are occupied in the wave function of the universe and, since electrons are fermions, a state cannot be occupied twice. Any unoccupied state of negative energy then appears like a positron of positive energy.) Some authors \cite{CS07,DEO16} have proposed that electrons are real whereas positrons are literally just empty spots in a sea of electrons of negative energy. Others \cite{crlet} have suggested that the situation is fundamentally symmetric between electrons and positrons, so that positrons are real particles in their own right, and that we are not surrounded by a sea of electrons that we usually do not notice; in such models, the Dirac sea has the status of a metaphor or mathematical analogy, but not of a reality.

\section{Philosophical Questions}

Bohmian mechanics is the subject of ongoing philosophical debate. One question concerns the ontological status of the wave function in Bohmian mechanics: Can a function on configuration space be physically real? Or is the wave function to be regarded as nomological, i.e., as something like a law (roughly analogous to the Hamiltonian function in classical mechanics), instead of as a thing \cite{GZ11}? Another question concerns the possible choices in quantum field theory, such as field ontology versus particle ontology: Which variant of the theory is the most convincing one? In some examples, different variants of Bohmian mechanics are empirically equivalent, so that no experiment can decide between them; it may nevertheless be possible, even natural, to prefer one over the other on theoretical or philosophical grounds (see \cite{aapr} for examples).

Another circle of philosophical questions concerns the following. Bohmian mechanics contains certain elements in its ontology, the particles with positions $\vQ_k(t)$, that immediately represent the distribution of matter in space. Such variables, and such elements of the ontology, are often called the \emph{primitive ontology} \cite{DGZ92,AGTZ14,Mau16}. I feel that for a fundamental physical theory to be acceptable, it must have a primitive ontology. This leads to questions such as: What exactly is required of a theory's ontology to make it acceptable as a fundamental physical theory? What kinds of things can constitute a primitive ontology? For example, Everett's many-worlds interpretation proposes that there is only wave function in the world; this does not seem to me to make enough sense as a fundamental physical theory (see also \cite{Mau10}), but this problem can be taken care of by introducing a suitable primitive ontology \cite{AGTZ11}, which leads to a modified theory with many-worlds character that makes clear sense.

\section{Further Reading}

A classic article is Bell's \cite{Bell86b} overview of different interpretations of quantum mechanics. 
An introductory article about Bohmian mechanics, particularly for philosophers, can be found in \cite{Gol01}.
Book-length introductions are provided in \cite{DT,Bri16}. Physical and mathematical applications of Bohmian mechanics are discussed in the collections \cite{Chat,Ori}.

\bigskip

\noindent{\it Acknowledgment.} I thank Travis Norsen and the editors for comments on a draft of this article.

\end{document}